\documentclass[8pt,aps,prl,nofootinbib,twocolumn,latexsym]{revtex4}
\usepackage{amssymb}

\usepackage{amsmath}
\usepackage[dvips]{graphicx}
\usepackage{subfigure}

\begin{document}

\title{Metal-Insulator Transition in a System of Superconducting Vortices Caused by a Metallic Gate.}
\author{K.~Michaeli and A.~M.~Finkel'stein}

\begin{abstract}
We address a recent experiment in which a strong decrease of the resistance of a
superconducting film has been observed when a remote unbiased gate was placed above
the film. Here we explain the experimental finding as a suppression of the vortex
tunneling due to the Orthogonality Catastrophe of the electrons inside the gate. We
interpret the change in the resistance of the film as a "metal-insulator" transition
in the system of vortices induced by the gate.
\end{abstract}

\affiliation{Department of Condensed Matter Physics, The Weizmann Institute of Science,
Rehovot 76100, Israel}
\maketitle


The dissipationless flow of electrons, which is the hallmark of superconductivity,
can be destroyed in the presence of a magnetic field by the motion of
vortices~\cite{Bardeen1965}. The dissipation is caused by nonsuperconducting
electrons located inside the vortex core. Paradoxically, the superconducting
properties can be restored by increasing the amount of disorder because
imperfections create a pinning potential~\cite{Larkin1970, Larkin1972} that opposes
the motion of vortices. The vortices may still hop
between the minima of the pinning potential by thermal activation~\cite%
{AndersonKim}, or at low temperatures by quantum tunneling~\cite%
{Glazman1992,Ephron1996,Mooij1996,Kogan2005}. Therefore, the observation of
a strong decrease of the resistance when an unbiased metallic gate is placed
above an amorphous superconducting film~\cite{Mason2002} is of great
interest.

Here we show how the response of electrons inside the metallic gate to a
change in the vortex positions can suppress the tunneling of the vortices,
thereby reducing the resistance of the film. We attribute the change in the
resistance to a magnetic coupling between the film and the gate. The
magnetic flux of a vortex penetrating inside the gate scatters the electrons
in a way similar to Aharonov-Bohm (\textit{A-B}) scattering~\cite%
{Aharonov1959,Aharonov1984}. The response of the electrons to a sudden
change in the vortex positions leads to the Orthogonality Catastrophe (%
\textit{OC}) that manifests itself in the vanishing overlap $\langle \Psi
_{f}|\Psi _{i}\rangle $ of the two wave functions describing the macroscopic
electron system before and after the change in the scattering potential~\cite%
{AndersonOC1967}. The tunneling vortices force the electrons inside the gate to
adjust themselves to the change in the vortex positions. In response, the electrons
oppose the tunneling and can even localize the vortices restoring the perfect
superconducting properties of the film at low temperatures. We interpret the
experiment~\cite{Mason2002} as a ''metal-insulator'' transition in a system of
tunneling vortices induced by the gate.

\begin{figure}[bp]
\begin{flushright}\begin{minipage}{.5\textwidth}  \centering \subfigure[]{
        \label{fig:metal} 
        \includegraphics[width=0.45\textwidth]{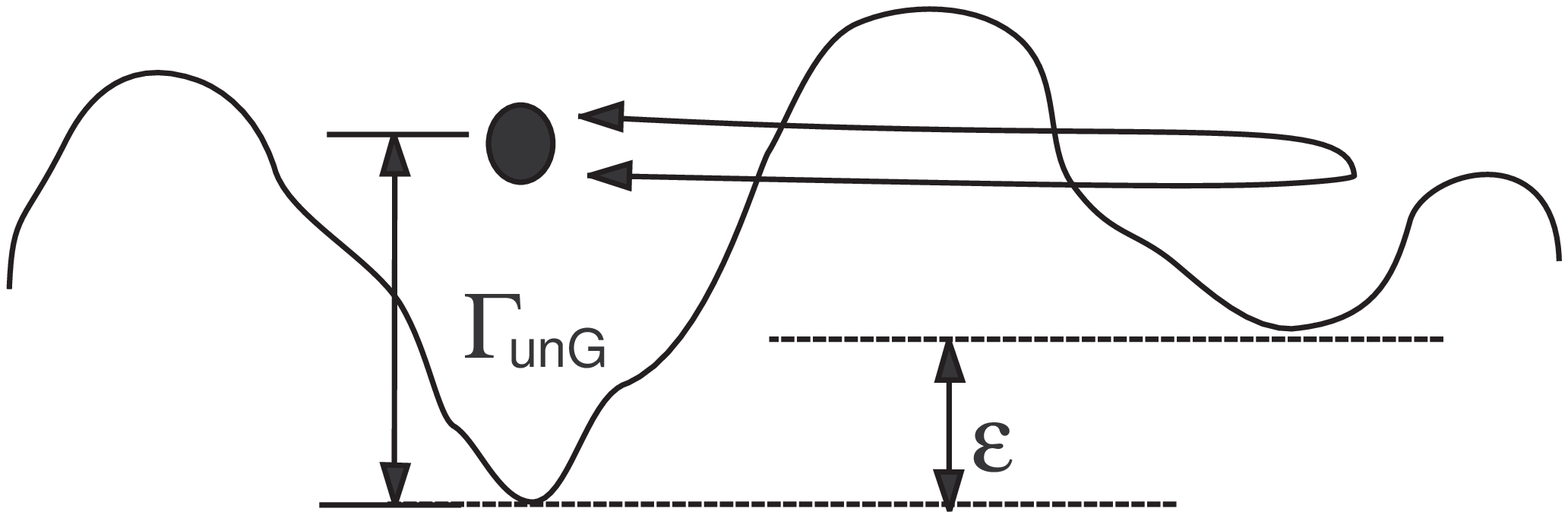}} \hspace{0.05in}%
        \subfigure[]{
        \label{fig:insulator} 
        \includegraphics[width=0.45\textwidth]{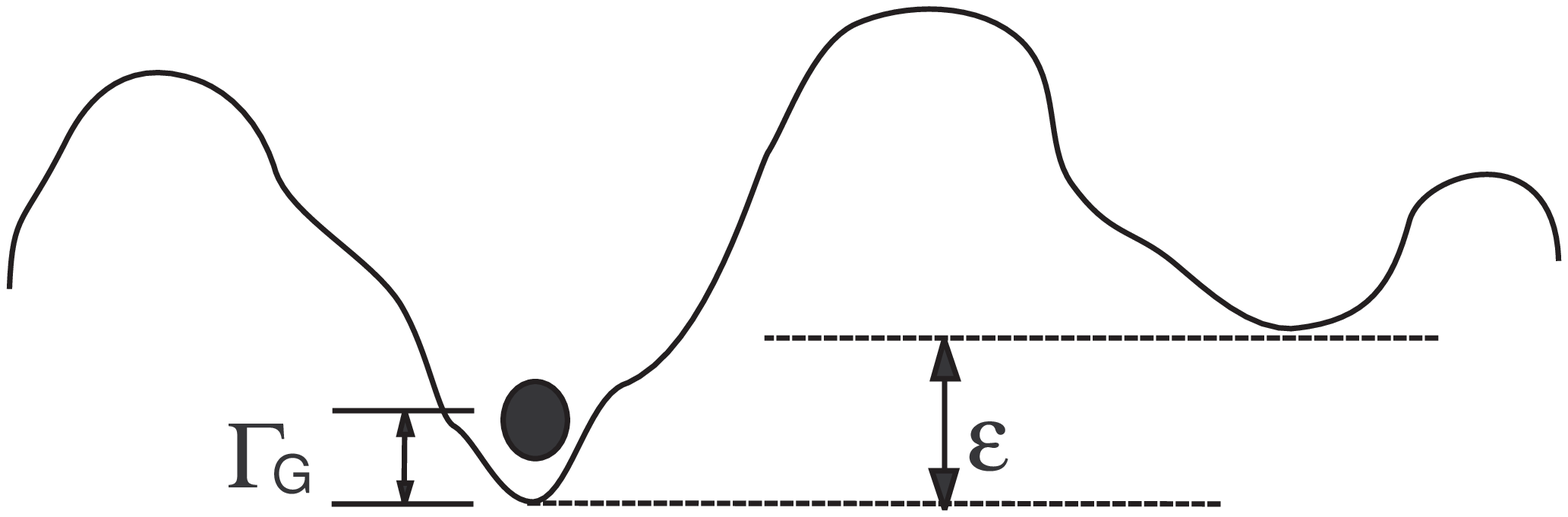}}
          \caption[0.4\textwidth]{\small The "metal-insulator"
          transition. The vortex in the effective potential
          landscape is represented by a hopping "particle".
        (a) The tunneling rate in the absence of the gate $\Gamma_{unG}$ exceeds the energy difference between the potential minima $\varepsilon$; the vortices are in a "metallic" phase
        (b) The tunneling rate is reduced by the gate to $\Gamma_{G}<\varepsilon$, and the system of vortices becomes an "insulator"; the dissipationless property of the superconducting film is restored.} \label{fig:MI}
\end{minipage}\end{flushright}
\end{figure}

Little is known about the quantum motion of the vortex density at low
temperatures~\cite{Fisher1991}. Fortunately, for studying the role of the
gate it is sufficient to assume that the change in the vortex positions is a
rare tunneling event. This can be tunneling of a single vortex, a bundle of
vortices, or topological defects such as dislocation pairs in the case of a
vortex lattice. Phenomenologically, the change in the vortex positions can
be described by the hopping Hamiltonian
\begin{equation}
H=\sum_{i}\varepsilon _{i}a_{i}^{+}a_{i}+\sum_{\left\langle i,j\right\rangle
}\left( \Gamma _{ij}a_{i}^{+}a_{j}+h.c.\right) .
\label{eq: Quantum Tunneling}
\end{equation}%
According to the standard criterion of the metal-insulator transition~\cite%
{Anderson1958} the ratio of the variation of the potential minima $%
\varepsilon =\left\langle \varepsilon _{i}\right\rangle $ to the typical
value of the tunneling rates $\Gamma =\left\langle \Gamma _{ij}\right\rangle
$ specifies whether vortices are itinerant or localized. The finite
resistance at low temperature in the absence of the gate indicates that the
vortices are mobile, i.e., the tunneling rate in the ungated film $\Gamma
_{unG}>\varepsilon $. The dissipationless nature of the superconducting film
is revived when the tunneling rate is reduced by the gate to $\Gamma
_{G}<\varepsilon $; see Fig.~\ref{fig:MI} which illustrates the two cases.

Following the experiment~\cite{Mason2002}, we assume that the gate does not
change the superconducting properties (such as $T_{c}$), but rather affects
only the motion of vortices by renormalizing the tunneling rate:%
\begin{equation}
\Gamma _{G}=\Gamma _{unG}\langle \Psi _{f}|\Psi _{i}\rangle .
\label{eq:Overlap1}
\end{equation}%
The overlap $\langle \Psi _{f}|\Psi _{i}\rangle $ accounting for the response of the
electron gas can be analyzed in terms of low energy electron-hole excitations
''decorating'' the vortex tunneling. One may consider the cloud of virtual
excitations as part of the tunneling process that lasts long after the change in the
vortex positions occurs; as a result of the substantial increase of the overall time
of motion along the tunneling trajectory, the tunneling rate is suppressed.
Formally, the overlap factor in Eq.~(\ref{eq:Overlap1}) can be presented in terms of
the operators $\hat{S}_{i}$ and $\hat{S}_{f}$ describing the scattering of the
electrons by the magnetic field of the tunneling vortices in their initial and final
positions~\cite{Yamada1982}:
\begin{equation}
|\langle \Psi _{f}|\Psi _{i}\rangle |=N^{-K};\hspace{1mm}K=-\frac{1}{8\pi
^{2}}Tr\left\{ \ln ^{2}(\hat{S}_{f}\hat{S}_{i}^{-1})\right\} .
\label{eq:Overlap_tunneling}
\end{equation}%
Here, $N$ is the number of electrons and hence the overlap factor vanishes
unless there is a mechanism that limits the effectiveness of the \textit{OC}%
. At finite temperatures~\cite{Yamada1984,Kagan1986} the large parameter $N$
should be substituted by another large parameter $(\max \{T,\Gamma
_{G}\}\tau _{tun})^{-1}$. Here $\tau _{tun}$ is the time of vortex tunneling
in the absence of the gate; $\tau _{tun}^{-1}$ acts as the high-energy
cutoff because only slow excitations that cannot follow adiabatically the
tunneling particle reduce the overlap factor. On the other hand, the
temperature enters as a low-energy cutoff because the excitations with
energy smaller than $T$ being thermally activated do not contribute to the
orthogonality of the initial and final states. In addition, electrons with
energies smaller than $\Gamma _{G}$ cannot react to the tunneling events as
they are too frequent for them. This is why the tunneling rate determines
its own renormalization in a self-consistent way:
\begin{equation}
\Gamma _{G}(T)=\Gamma _{unG}(\max \{T,\Gamma _{G}\}\tau _{tun})^{K}.
\label{eq:Gamma}
\end{equation}%
For $K<1,$ the tunneling rate remains finite at low temperatures $\Gamma
_{G}(T\rightarrow 0)=\Gamma _{unG}(\Gamma _{unG}\tau _{tun})^{K/(1-K)}$,
while for $K>1$ the tunneling rate $\Gamma _{G}(T)$ goes to zero with the
temperature as $T^{K}$.

\begin{figure}[]
\begin{flushright}\begin{minipage}{0.5\textwidth} \centerline{
    \includegraphics[bb={73 433 541 770},width=0.9\textwidth]{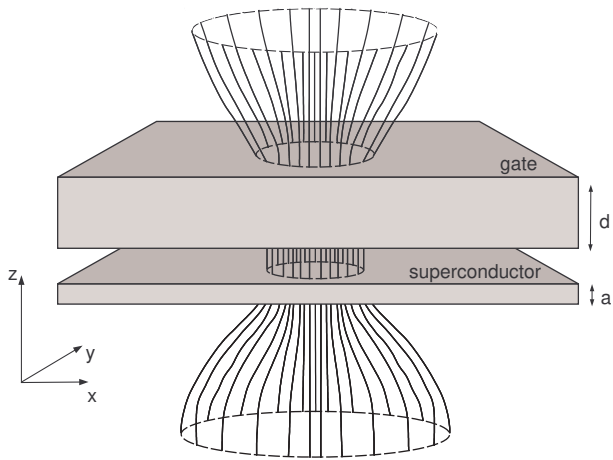}}
\caption{\small Superconducting film magnetically
    coupled to a metallic gate. The magnetic field of the
vortex penetrates the gate.} \label{fig:vortex_field}
\end{minipage}\end{flushright}
\end{figure}
The renormalization of $\Gamma _{G}$ induced by the gate can explain the
localization of vortices (i.e. a transition from $\Gamma _{unG}>\varepsilon $ to
$\Gamma _{G}<\varepsilon $) if the exponent $K$ is comparable to or larger than one.
In the following part of this Letter we show that for a
superconducting film magnetically coupled to a metallic gate (see Fig.~\ref%
{fig:vortex_field}) the exponent $K$ is
\begin{equation}
K=\varsigma (\alpha \delta r)^{2}\frac{(k_{F}^{gate})^{2}d}{64R}.
\label{eq:K}
\end{equation}%
Here $\delta r$ is the typical distance that vortices have to tunnel, which
is approximately the coherence length~\cite{Larkin1972}, $\delta r\sim \xi $%
. The parameter $\alpha $ is the total flux measured in units $\Phi
_{0}=2\pi \hbar c/e$ that moves a distance $\delta r$ as a result of a
tunneling event. For a single vortex, $\alpha =1/2$; the same holds for
interstitials or vacancies. When a bundle (or a pair of dislocations) is
tunneling, $\alpha $ should be multiplied by the number of vortices. The
result is universal and valid as far as $\delta r\ll R$, where $R$ is the
radius of the area inside the gate occupied by the magnetic field of a
vortex, which in the present geometry~\cite{Mason2002} is about the
superconducting penetration depth, $R\approx \lambda $. The expression for $%
\lambda $ is determined by the vortex solution specific for thin-film
superconductors, known as the Pearl vortex~\cite{Pearl1964, Abrikosov}.
Other factors determining $K$ are the gate thickness $d$, and the Fermi
momentum of the electrons in the gate $k_{F}^{gate}$. The prefactor $%
\varsigma $ is evaluated numerically as $\approx 0.4$.

Using the expressions for $\lambda $ and $\xi$ in disordered thin films the
exponent can be rewritten as:
\begin{equation}
K\sim \varsigma \frac{\alpha ^{2}}{48\pi }\left( \frac{e^{2}}{{c}}\right)
^{2}\frac{v_{F}^{sc}}{e^{2}}%
(k_{F}^{gate}d)(k_{F}^{gate}a)(k_{F}^{sc}l^{sc})^{2}.  \label{eq:K-general}
\end{equation}%
The index $sc$ refers to the electrons in the superconducting film: $l^{sc}$
is their mean free path (in the normal state) and $v_{F}^{sc}$ is the Fermi
velocity; $a$ is the thickness of the film, and $\hbar =1$. Interestingly, $%
T_{c}$ drops out from $K$ so that it depends only on the geometrical factors and the
nonsuperconducting properties of electrons. We see that the value of the exponent
$K$ is determined by a small factor $\sim 10^{-7}$ opposed by a product of a few
large factors. The condition for the vortex localization can be easily fulfilled for
a not too thin gate and not too disordered superconducting film.

Next, we briefly sketch the steps in the derivation of Eq.~(\ref{eq:K})
starting from Eq.~(\ref{eq:Overlap_tunneling}). Let us consider the \textit{%
OC} in response to a single vortex tunneling. The cylindrical symmetry of
the vortex allows us to analyze the scattering of electrons using the basis
of cylindrical waves, $|\ell ,q,k_{z}\rangle $; here $\ell $ is the angular
momentum along the $z$ axis, while $q$ and $k_{z}$ are the magnitudes of the
in-plane and $z$ components of the momentum. In this basis the scattering
operator is diagonal and can be described in terms of the phase shifts
\begin{equation}
\langle {\ell ,q,k_{z}}|S|\ell ^{\prime },q^{\prime },k_{z}^{\prime }\rangle
=e^{2i\delta _{\ell }}\delta _{\ell ,\ell ^{\prime }}\delta _{q,q^{\prime
}}\delta _{k_{z},k_{z}^{\prime }}.
\label{eq:Phase_Shift_Cylindrical_Coordinates}
\end{equation}%
Since the exponent in Eq.~(\ref{eq:Overlap_tunneling}) contains a product of
two scattering operators corresponding to the different vortex positions
shifted by $\delta r$, we have to use the transformation matrix between the
two\emph{\ }bases of cylindrical waves centered at these positions: $%
_{f}\langle {\ell ,q,k_{z}}|{\ell ^{\prime }q^{\prime },k_{z}^{\prime }}%
\rangle _{i}=J_{(\ell -\ell ^{\prime })}(q\delta r)\delta _{q,q^{\prime
}}\delta _{k_{z},k_{z}^{\prime }}$, where $J_{\nu }(z)$ is the Bessel
function. The elements of the matrix $S_{f}S_{i}^{-1}$ can be easily
calculated as
\begin{equation}
_{f}\langle {\ell }|S_{f}S_{i}^{-1}|{\ell ^{\prime }}\rangle
_{f}=\sum_{n}e^{2i\delta _{\ell }-2i\delta _{n+\ell }}J_{n}(q\delta
r)J_{n-\ell ^{\prime }+\ell }(q\delta r)\ .
\label{eq:Overlap_Matrix_Element}
\end{equation}

To proceed further, we need to find the phase shifts specific for the
scattering by the vortex. An analogy to classical scattering, where the
angular momentum is related to the impact parameter $b=|\ell |/q,$ helps
elucidate the behavior of the phase shift as a function of $\ell $. For $%
b\gg {R},$ the scattering by the vortex is similar to the \textit{A-B}
scattering by a flux $\alpha \Phi _{0}$. In the \textit{A-B} scattering~\cite%
{Aharonov1959,Aharonov1984} electrons acquire the phase $\delta _{\ell
}^{A-B}=\frac{\pi }{2}(|\ell |-|\ell -\alpha |)$. The uniqueness of this
scattering is in its infinite range: $\delta _{\ell }$ does not vanish when $%
|\ell |\rightarrow \infty $. For scattering by the vortex, the jump in the
\textit{A-B} phase shifts is smeared out, but the infinite range character
of this scattering is preserved. Hence, $\delta _{\ell }$ varies
monotonically as a function of $\ell $ between the two limits:%
\begin{equation}
\delta _{\ell }\xrightarrow[\ell{\gg}qR]{}\alpha \frac{\pi }{2}sgn\ \ell .
\label{eq:AB_Phase_Shift}
\end{equation}%
Naturally, for $qR\gg 1$ the phase shift depends on $b$ and $R$ only through
the dimensionless combination $b/R=\ell /qR$ such that $\delta _{\ell }=%
\frac{\alpha \pi }{2}g\left( \ell /qR\right) $; see Fig.~\ref%
{fig:Finite_Core-Phase_Shift} for illustration.

\begin{figure}[h]
\begin{flushright}\begin{minipage}{0.5\textwidth} \centerline{
   \includegraphics[bb= {0 0 549.12 403.2}, width=0.95\textwidth, height=0.65\textwidth]{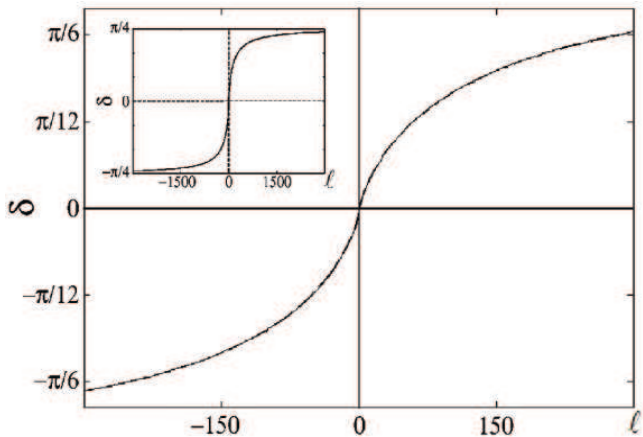}}
\caption{\small The phase shift for an electron scattered by the magnetic field of a
superconducting vortex as a function of the angular momentum. The phase shift is
calculated for $qR$=50, 100, and 150. The three curves are rescaled to $qR$=100 to
demonstrate the universality of the scattering. The insert shows the asymptotic
behavior of the phase shift at large angular momenta,
$\delta_{\ell}\rightarrow\pm{\pi}/{4}$ .} \label{fig:Finite_Core-Phase_Shift}
\end{minipage}\end{flushright}
\end{figure}

We now notice that the sum determining the elements of $_{f}\langle {\ell }%
|S_{f}S_{i}^{-1}|{\ell ^{\prime }}\rangle _{f}$ is accumulated at $-q\delta
r\lesssim n\lesssim q\delta r$. This is because the Bessel Functions $J_{\nu
}(z)$ decay exponentially with their order when $\nu >z$. Therefore, since
for thin superconductors $\delta r/R\sim \xi /\lambda \ll 1$, the phase
shifts difference in Eq.~(\ref{eq:Overlap_Matrix_Element}) can be
approximated as:
\begin{equation}
\delta _{\ell }-\delta _{n+\ell }\xrightarrow[n{\ll}qR]{}-n\delta _{\ell
}^{\prime };\qquad \delta _{\ell }^{\prime }\approx \frac{\alpha \pi }{2qR}%
g^{\prime }\left( \frac{\ell }{qR}\right) \ll 1.
\label{eq:Phase_shifts_Difference}
\end{equation}%
The final step of the calculation is to expand in $\delta r/R$ the logarithm
in Eq.~(\ref{eq:Overlap_tunneling}), and take the trace over $\ell $ and the
momentum on the Fermi surface. The outcome of the calculation is given in
Eq.~(\ref{eq:K}). The gate thickness $d$ appears here as a result of taking
the trace. The specifics of the vortex solution enter only through $g(x)$,
with the integral yielding $\varsigma =\int {dx(dg/dx)}^{2}\approx 0.4$.

The expression in Eq.~(\ref{eq:K}) can be applied for any bundle with a
total flux $\alpha \Phi _{0}$ that moves a distance $\delta r$ as a result
of a tunneling event. This is because the magnetic field of the tunneling
vortices extends over a large distance, so that their exact configurations
before and after the tunneling are not important. The only relevant quantity
is the product $\alpha \delta r$.

Finally, note that although we invoke the expansion in terms of $\delta
r/R\ll 1,$ we get $K\propto (\delta r)^{2}/R.$ This is a typical feature of
the \textit{OC} in the case of an extended scattering potential~\cite%
{MatveevLarkin1992} when a large number of harmonics is involved. This can
be understood from the following arguments. It has been shown that the
\textit{OC} in the discussed problem is determined by $\sum_{\ell }(\delta
_{\ell }-\delta _{\ell +1})^{2}\approx \sum_{\ell }(\delta _{\ell }^{\prime
})^{2}$. Since the phase shifts approach asymptotically the limit $\pm
\alpha \frac{\pi }{2},$ then the sum
\begin{equation}
\sum_{\ell }(\delta _{\ell }-\delta _{\ell +1})\approx \sum_{\ell }\delta
_{\ell }^{\prime }=\pi \alpha .  \label{eq:Sum_of_differnces}
\end{equation}%
Therefore, the result obtained for the exponent $K$ corresponds to the
differences $(\delta _{\ell }-\delta _{\ell +1})$ that are distributed
almost equally between $L\sim qR$ channels:
\begin{equation}
\sum_{\ell }(\delta _{\ell }-\delta _{\ell +1})^{2}\sim L\left( \frac{\pi
\alpha }{L}\right) ^{2}\sim \frac{(\pi \alpha )^{2}}{qR}.
\label{eq:Sum_of_squares}
\end{equation}%
The above arguments helps to understand the effect of disorder in the gate
on the \textit{OC}. One may expect that the randomization of the phase
shifts due to the disorder can only increase the value of the exponent $K.$
In the general case, $\ell $ should be substituted by an index $i$ of the
scattering channel (i.e., the index of the states diagonalizing the
scattering matrix ). The scattering by impurities leads to the randomization
of the phase differences, while the asymptotic limits of the phase shifts
remain the same, $\pm \alpha \frac{\pi }{2}$. Therefore, the value of the
exponent $K$, which is determined by the squares of the phase differences,
should increase in the presence of disorder. [Under the condition~(\ref%
{eq:Sum_of_differnces}), the obtained exponent $K\propto (\delta r)^{2}/R$
is close to the minimum possible value which is at equal phase differences.]
Our conclusion that disorder increases the effect of the gate in suppressing
the tunneling rate $\Gamma _{G}$ is in accordance with the existing
theoretical results about the enhancement of the \textit{OC-}exponent by not
too strong disorder~\cite{Kroha1992, Gefen2002}.

The idea to use a double layer system to study dynamics of vortices is well
known~\cite{Giaever1965, Klapwijk1991, Rimberg1997}. The peculiarity of the
discussed experiment~\cite{Mason2002} is that the resistance of a superconducting
film has been measured at various magnetic fields both with and without the gate. In
the absence of the gate, the resistance initially decreases with lowering the
temperature but eventually saturates at finite values, strongly supporting the
possibility of vortex tunneling. On the other hand, when the film is gated the
resistance drops with no indication of saturation. The two behaviors begin to
deviate at the same temperatures at which the saturation of the resistance in the
ungated film occurs. In other words, the gate affects the vortex motion in the
tunneling regime only. This supports our claim that the gate reduces the tunneling
rate of the vortices. (The gate is not effective in slowing down a continuous flow
of vortices as the electrons can follow the vortices adiabatically.) We interpret
the marked difference in the low temperature behavior of the film with and without
the gate as a ''metal-insulator'' transition in a system of tunneling vortices
induced by the gate.

In addition to the magnetic coupling between the film and the gate, one may
consider a capacitive coupling between them. In the case of the Josephson
junction arrays (or granular superconductors) the capacitive coupling
reduces the fluctuations of the phase of the superconducting order parameter~%
\cite{Rimberg1997, Wagenblast1997}. As a result, the system may undergo a
transition from an insulating to a superconducting state. However, for a
homogenous film with a relatively small resistance $\sim 1.5\ k\Omega
/\square $ used in Ref.~\cite{Mason2002} the phase fluctuations are not so
effective~\cite{Ramakrishnan1989}. This is confirmed by the observed
insensitivity of the critical magnetic field $H_{c}$ to the presence of the
gate. Furthermore, in homogeneous superconductors the motion of vortices is
not accompanied by the redistribution of the charge density. Therefore,
there are good reasons to ignore the capacitive coupling between the film
and the gate.

The Eddy currents (Foucault currents) generated inside the gate by the emf
as a result of the magnetic coupling can also contribute to the suppression
of $\Gamma _{G}$. We find that under the conditions of the experiment~\cite%
{Mason2002} the \textit{OC} is dominant. Experimentally, one can identify
the main mechanism of the gate response to the vortex tunneling by changing
the conductivity of the gate.

In conclusion, vortices are in the core of any physical picture describing
the quantum phase transitions in superconducting films. The gated system
discussed here can be used as an effective tool for investigating the
microscopics of the vortex motion at low temperatures. It provides a unique
opportunity to study the vortex tunneling in thin superconducting films by
such simple means as varying the characteristics of the gate, in particular
the gate thickness.

\begin{acknowledgments}
We thank A.~Kapitulnik, T.~M.~Klapwijk and B.~Spivak for useful discussions.
AF is supported by the Minerva Foundation.
\end{acknowledgments}


\begin{thebibliography}{99}
\bibitem{Bardeen1965} J.~Bardeen, and M.~J.~Stephen, \textit{Phys.~Rev.}~%
\textbf{140}, 1197 (1965).

\bibitem{Larkin1970} A.~I.~Larkin, \textit{Zh.~Eksp.~Teor.~Fiz. }~\textbf{58}%
, 1466 (1970) [\textit{Sov.~Phys.~JETP.}~\textbf{31}, 784 (1970)].

\bibitem{Larkin1972} A.~I.~Larkin, and Yu.~N.~Ovchinnikov, \textit{%
Zh.~Eksp.~Teor.~Fiz.}~\textbf{61}, 1221 (1971) [\textit{Sov.~Phys.~JETP.}~%
\textbf{34}, 651 (1972)].

\bibitem{AndersonKim} P.~W.~Anderson, and Y.~B.~Kim, \textit{Rev.~Mod.~Phys.}%
~ \textbf{36}, 39 (1964).

\bibitem{Glazman1992} Y.~Liu, D.~B.~Haviland, L.~I.~Glazman, and
A.~M.~Goldman, \textit{Phys.~Rev.~Lett.}~\textbf{68}, 2224 (1992).

\bibitem{Ephron1996} D.~Ephron, A.~Yazdani, A.~Kapitulnik, and
M.~R.~Beasley, \textit{Phys. Rev. Lett.} \textbf{76}, 1529 (1996).

\bibitem{Mooij1996} A.~van~Oudenaarden, S.~J.~K.~Vardy, and J.~E.~Mooij,
\textit{Phys. Rev. Lett.}~\textbf{77}, 4257 (1996).

\bibitem{Kogan2005} F.~Tafuri et al., \textit{Europhys. Lett.}~\textbf{73},
948 (2006).

\bibitem{Mason2002} N.~Mason, and A.~Kapitulnik, \textit{\ Phys.~Rev.~B.}~%
\textbf{65}, 220505(R) (2002).

\bibitem{Aharonov1959} Y.~Aharonov, and D.~Bohm, \textit{Phys.~Rev.}~\textbf{%
115}, 485 (1959).

\bibitem{Aharonov1984} Y.~Aharonov, C.~K.~Au, E.~C.~Lerner, and J.~Q.~Liang,
\textit{Phys.~Rev.~D}~\textbf{29}, 2396 (1984).

\bibitem{AndersonOC1967} P.~W.~Anderson, \textit{Phys.~Rev.~Lett.}~\textbf{18%
}, 1049 (1967).

\bibitem{Fisher1991} The quantum tunneling of vortices leading to the
variable-range hopping resistivity in superconducting films has been
discussed by M.~P.~A.~Fisher, T.~A.~Tokuyasu, and A.~P.~Young, \textit{Phys.
Rev. Lett.}~\textbf{66}, 2931 (1991).

\bibitem{Anderson1958} P.~W.~Anderson, \textit{Phys.~Rev.}~\textbf{109},
1492 (1958).

\bibitem{Yamada1982} K.~Yamada, and K.~Yosida, \textit{Prog.~Theor.~Phys.}~%
\textbf{68}, 1504 (1982).

\bibitem{Yamada1984} K.~Yamada, \textit{Prog.~Theor.~Phys.}~\textbf{72}, 195
(1984).

\bibitem{Kagan1986} Yu.~Kagan, and N.~V.~Prokof'ev, \textit{%
Zh.~Eksp.~Teor.~Fiz.}~\textbf{90}, 2176 (1986) [\textit{Sov.~Phys.~JETP}~%
\textbf{63}, 1276 (1986)].

\bibitem{Pearl1964} J.~Pearl, \textit{App.~Rev.~Lett.}~\textbf{5}, 65
(1964).

\bibitem{Abrikosov} A.~A.~Abrikosov, \textit{Fundamentals of the Theory of
Metals} (Elsevier, Amsterdam, 1988).

\bibitem{MatveevLarkin1992} K.~A.~Matveev, and A.~I.~Larkin, \textit{Phys.
Rev. B}~\textbf{46}, 15337 (1992).

\bibitem{Kroha1992} Y.~Chen, and J.~Kroha, \textit{Phys.~Rev.~B.}~\textbf{46}%
, 1332 (1992).

\bibitem{Gefen2002} Y.~Gefen, R.~Berkovits, I.~V.~Lerner and
B.~L.~Altshuler, \textit{Phys.~Rev.~B.}~\textbf{65}, 081106(R) (2002).

\bibitem{Giaever1965} I.~Giaever, \textit{Phys.~Rev.~Lett.}~\textbf{15}, 825
(1964).

\bibitem{Klapwijk1991} G.~H.~Kruithof, P.~C.~van~Son, and T.~M.~Klapwijk,
\textit{Phys.~Rev.~Lett.}~\textbf{67}, 2725 (1991).

\bibitem{Rimberg1997} A.~J.~Rimberg et al., \textit{Phys.~Rev.~Lett.}~
\textbf{78}, 2632 (1997).

\bibitem{Wagenblast1997} K.~H.~Wagenblast, A.~van~Otterlo, G.~Schon, and
G.~T.~Zimanyi, \textit{Phys.~Rev.~Lett.}~\textbf{79}, 2730 (1997).

\bibitem{Ramakrishnan1989} T.~V.~Ramakrishnan, \textit{Physica~Scripta.}~
\textbf{T27}, 24 (1989).
\end{thebibliography}
\end{document}